\documentclass{article}              

\usepackage{graphicx}          
\usepackage{amssymb,amsmath}
\usepackage[noend]{algpseudocode}
\usepackage{xcolor}
\usepackage[linesnumbered,ruled,vlined]{algorithm2e}

\DeclareMathOperator{\E}{\mathrm{E}}
\newcommand{{\Cov}}{\mathrm{Cov}}
\newcommand{{\Var}}{\mathrm{Var}}
\newcommand*{\defeq}{\stackrel{\text{def}}{=}}

\SetCommentSty{mycommfont}
\SetKwInput{KwInput}{Input}                
\SetKwInput{KwOutput}{Output}              

\newcommand\smallO{
  \mathchoice
    {{\scriptstyle\mathcal{O}}}
    {{\scriptstyle\mathcal{O}}}
    {{\scriptscriptstyle\mathcal{O}}}
    {\scalebox{.7}{$\scriptscriptstyle\mathcal{O}$}}
  }

\title{A simple and robust method for noise variance estimation for time-varying signals}
\author{Qin Li \\
	Equinor Research Centre, Porsgrunn, 3936 Norway  \\
	\and
	Junchan Zhao \\
	School of Mathematics and Statistics \\
    Hunan Univ. of Technology and Business \\
    Changsha, 410205 China \\
	}

\date{\today}
\begin{document}

\maketitle

\begin{abstract}
In this brief paper, we present a simple approach to estimate the variance of measurement noise with time-varying 1-D signals. The proposed approach exploits the relationship between the noise variance and the variance of the prediction errors (or innovation sequence) of a linear estimator, the idea that was pioneered by \cite{ref:Mehra-1970} and \cite{ref:Belanger-1974}. Compared with the classic and more recent methods in the same category, the proposed method can render more robust estimates with the presence of unmodelled dynamics and outliers in the measurement.
\end{abstract}

\section{Introduction}
In many industrial processes, one frequently faces a measured data that is composed of the true value of a signal and some random noise. One central problem in signal processing is to find the characteristics of noise, which is critical to many important tasks such as signal filtering, estimation, prediction and anomaly detection. The probabilistic variance is among the most interesting characters of noise, hence the estimation of it has been a popular topic in many research fields.

In image or speech processing, the estimation of noise properties is mainly linked to denoising processes; several methods have been proposed which deal with the data either in its original space \cite{ref:Immerker-1996} or represented in some transformed (e.g. via FFT and wavelets) basis \cite{ref:Sijbers-2007}, \cite{ref:Hashemi-2010,ref:Beheshti-2003}. In systems and controls society, the estimation of noise properties have been important for signal estimation and system identification and attracted a great deal of research interest in 1970s and 1980s. Various methods of different types were proposed in that time. Among them, two groups of methods have been reviewed and improved in the past decades, namely the correlation methods and joint state and parameter estimation methods. The paper \cite{ref:Dunik-2009} give an brief yet extensive survey of the previous works and shows a simulation-based comparison with selected methods. The result of the comparison favours the correlation methods that was pioneered by the works in \cite{ref:Mehra-1970,ref:Belanger-1974} and have been improved and extended by \cite{ref:Odelson-2006,ref:Dunik-2008,ref:Ge-2014,ref:Dunik-2018}. The basic idea of the correlation method is to link the auto-correlations of the prediction errors of a sub-optimal linear estimator to the variances of the noises in the state-space equations describing the system. Least-squares fitting is then used to find the best matching variances of the noises. In principle, the method needs the knowledge of the deterministic part of the system and the noise properties are assumed time-invariant; hence its performance is not warranted with unknown system dynamics, inputs, disturbances and time-varying noise properties (The robust issue can be seen in the simulation in Section \ref{sec:sim}). The robustness is certainly important to industrial application yet has not been well examined in the literature.

In this paper, we follow the line of correlation method and propose a simple and robust method for noise variance estimation with time-varying signals. Unlike most methods existing in this category, our focus is laid only on the estimation of the variance of measurement (or sensor) noise. The time-varying signal under our consideration is 1-D and can be considered output of a system with unmodelled dynamics and disturbances (The detailed description of the signal and noise is given in Section \ref{sec:signal}). We present our method, which utilizes a robust measure of the variability of the innovations, in Section \ref{sec:sec-method} and give an analysis of its estimation error in \ref{sec:analysis}. In Section \ref{sec:sim} the proposed method is compared with two popular methods in the literature with a simulation example, where we show the robust issue of the existing methods and demonstrate how it is handled by the proposed method. We draw conclusions in Section \ref{sec:conclusions}.

\section{The signal and measurement} \label{sec:signal}
Consider a time-varying {\it signal} $\{x_k\}$, where $k=0,1,2,\cdots,$ is the discrete time indices. Suppose that we cannot measure the actual values of the signal but instead we have a sensor measurement in the form of $y_k = x_k + v_k$, where $v_k$ is a white noise sequence with
\begin{equation}\label{eq:meas-noise}
\E(v_k) = 0, \quad \Cov(v_k v_l) = R \delta_{kl}.
\end{equation}
for all $k,l=0,1,2,\cdots$. Our goal is to estimate the noise variance $R$ by using the measurement data \footnote{In practice, the noise variance often has event-based changes and thus is piece-wise constant. For simplicity we will not focus our synthesis and analysis on time-varying variances. But we will remark on the trackability of our estimator later in this section and demonstrate it in the simulation example in Section \ref{sec:sim}.}.

It is straightforward to see that the signal and measurement can be described by the following dynamical system:
\begin{subequations}\label{eq:sys}
\begin{align}
x_{k+1} &= x_k + w_k, \label{eq:sys-1} \\
y_k &= x_k + v_k, \label{eq:sys-2}
\end{align}
\end{subequations}
where $w_k$ is the change of the signal between time $k$ and time $k+1$. In this paper, we pursue an approach that is robust to unmodelled system dynamics and disturbances, hence we allow $w_k$ to take any form.

\section{Estimation method for measurement noise} \label{sec:sec-method}
In a conventional setting of the correlation based method for noise variance estimation (see, e.g., \cite{ref:Mehra-1970} and \cite{ref:Dunik-2009}), $w_k$ is assumed to be a sequence of independent random variables with identical distribution with zero mean value (That is, all other changes to the signal are separated out in the system equations). This means that $w_k$ satisfies that, for all $k,l=0,1,2,\cdots$,
\begin{equation}\label{eq:process-noise-mean-var}
\E(w_k) = 0, \quad \Cov(w_k w_l) = Q \delta_{kl}.
\end{equation}
In addition, $w_k$ is assumed uncorrelated with $v_k$, i.e., $\E\left(w_k v_l\right) = 0$.
A linear estimator for $x_k$, $k=1,2,\cdots$, can be designed as follows:
\begin{subequations}\label{eq:KF}
\begin{align}
\hat x^-_{k} &= \hat x_{k-1}, \label{eq:state-pred} \\
\hat x_{k} &= \hat x^-_{k} + K(y_{k} - \hat x^-_{k}).  \label{eq:state-update}
\end{align}
\end{subequations}
The state estimation error $e_k$ and the prediction error $\eta_k$ of the estimator is defined, respectively, as $e_k \defeq x_k - \hat x^-_k$ and $\eta_k \defeq y_{k} - \hat x^-_{k}$. The prediction error $\eta_k$ is usually called the \textit{innovation} (at time $k$). To have a stable estimator, we limit the value of $K$ to be inside $(0,1)$.

The key idea of the correlation method for noise variance estimation is utilizing the dependency of the covariance matrix of $e_k$ and auto-covariance of $\eta_k$ on the noise variance $Q$ and $R$. As derived in \cite{ref:Mehra-1970, ref:Odelson-2006}, the covariance matrix $\E(e_k e^\top_k)$ converges to
\begin{equation}\label{eq:ss-var-est-err}
  M = [M - K M - M K^\top + K(M+R)K^\top] + Q;
\end{equation}
and the covariance matrix $\E(\eta_k\eta^\top_k)$ converges to
\begin{align}\label{eq:ss-var-innovation}
  C & = M + R.
\end{align}
The matrix $M$ and $C$ are hence called the {\it steady-state} covariance matrix of the estimation error $e_k$ and the innovation $\eta_k$.

For 1-D signal, we can solve \eqref{eq:ss-var-est-err} and \eqref{eq:ss-var-innovation} to obtain an expression for the variance of measurement noise:
\begin{align}\label{eq:R}
  R & = \frac{C(2K-K^2)-Q}{2K}.
\end{align}
It is clear that one can obtain the value of $R$ with the knowledge of $C$ and $Q$ (as $K$ is up to our design). Now, one important choice we make to proceed with is to drop $Q$ in \eqref{eq:R}. This is because we want to handle the general situation in which the signal variation $w_k$ is unknown and not confined to \eqref{eq:process-noise-mean-var}. Ignoring $Q$ in \eqref{eq:R} we may approximate the noise variance simply as follows:
\begin{align}\label{eq:R-est}
  \hat R & = \frac{\hat C(2K-K^2)}{2K} = \hat C\bigg(1-\frac{K}{2}\bigg),
\end{align}
where $\hat C$ is an estimate of $C$, on which we will elaborate below.

Note that, in 1-D case, $C$ is just the steady-state variance of the innovation $\eta_k$. In practice, $\Var(\eta_k)$ normally converges fast and hence one may estimate $C$ using with a moving time window (which we will call a \textit{estimation window}). Let the size of the window be $m$, then a typical sample estimate of $C$ may be formed as
\begin{equation} \label{eq:C-est-mean}
\hat C(k,m) = \begin{cases}
\frac{1}{m}\sum_{i=k-m}^{k} \left(\eta_i - \left(\frac{1}{m+1}\sum_{j=k-m}^{k}\eta_j\right) \right)^2, & \text{if $k\geq m$}; \\
\frac{1}{k}\sum_{i=0}^{k} \left(\eta_i - \left(\frac{1}{k+1}\sum_{j=0}^{k}\eta_j\right) \right)^2, & \text{if $k < m$}.
\end{cases}
\end{equation}
The estimator in \eqref{eq:C-est-mean} is vulnerable to outliers in the measurement noise as well as the large abrupt changes in the signal, since both could generate large innovation that is {\it not} related to variance of the measurement noise. A much more robust estimator can be formed by utilizing the {\it sample median absolute deviation}:
\begin{equation} \label{eq:C-est-MAD}
\hat C(k,m) = \begin{cases}
\left(a \cdot \mathrm{med}\{|\eta_i - \mathrm{med}\{\eta_i: i\in\overline{k-m,k}\}|: i\in\overline{k-m,k}\}\right)^2, & \text{if $k\geq m$}; \\
\left(a \cdot \mathrm{med}\{|\eta_i - \mathrm{med}\{\eta_i: i\in\overline{0,k}\}|: i\in\overline{0,k}\}\right)^2, & \text{if $k < m$}.
\end{cases}
\end{equation}
where the notation ``$\mathrm{med}$" is the median operator for an finite set of real numbers and $\overline{m,n}$ denotes the integer sequence ${m, m+1, \cdots, n}$. The selection of parameter $a$ depends on the distribution of $\eta_k$. For example, if $\eta_k$'s are independent and follow a identical Gaussian distribution, then $a$ should be set as $1.4268$ and \eqref{eq:C-est-MAD} converges to \eqref{eq:C-est-mean} almost surely as $m$ goes to infinity (See \cite{ref:Pham-Gia-2001} for more properties about the median absolute deviations). In practice, it may be difficult to know the exact statistical distribution of $\eta_k$; it is probably then wise to make $a$ a tuning parameter.

It often happens in practice that the variance of measurement noise is varying with time or the magnitude of the signal, instead of being a constant value. To keep track the noise variance, one could compute $C(k,m)$ with progressing $k$. The window length $m$ should be chosen to balance the trackability and variability of the output of the estimator: With a long estimation window the estimate tends to be less fluctuating but lack the capability to track fast changes of the noise variance; on the other hand, with a short estimation window, the estimate may vary in a larger range but can respond to the change of noise variance quickly.

Summarizing what has been presented in this section, we propose the following algorithm for estimation of the variance of the measurement noise.

\vspace{0.5cm}

\begin{algorithm}[H] \label{alg:main}
\DontPrintSemicolon

  \SetKwInOut{Parameter}{parameter}
  \KwInput{Measurement sequence $y_k,k=0,1,2,\cdots$}
  \KwOutput{Estimated noise varaince at time $k$ (denoted by $\hat R_k$)}
  \Parameter{$K\in(0,1)$ (estimator gain), $m$ (window length)}
  \For{$k=0,1,2,\cdots$}
  {
    \If{$k=0$}
    {
        $\hat x_0 = y_0$;
    }
    \Else
    {      	
        $\hat x^-_{k} = \hat x_{k-1}$\;
        $\eta_k = y_{k} - \hat x^-_{k}$  \tcp*{innovation at step k}
        $\hat x_{k} = \hat x^-_{k} + K\eta_k$\;
        Estimate the innovation variance $\hat C$ by \eqref{eq:C-est-MAD}\;
        Estimate noise variance as $\hat R_k = \hat C\left(1-\frac{K}{2}\right)$
    }

  }

\caption{A simple noise variance estimation algorithm}
\end{algorithm}

We have mentioned the selection of the window length $m$ and will address the effect of the choice of $K$ in the next section, in which we give a informal analysis of the performance of the proposed approach.

\section{An informal analysis of the estimation error} \label{sec:analysis}
Naturally, we wish to quantify the estimation error with the proposed simple algorithm, which can be defined as $\epsilon(k,m)=\frac{2-K}{2}\hat C(k,m) - R$. We will show how this error is related to the variation of the signal. As the focus of the paper is a simple and practical estimation method, a rigorous mathematical proof is not pursued here. Furthermore, due to technical difficulties in dealing with median values, we only give an analysis for the estimation using Algorithm 1 with \eqref{eq:C-est-mean}, even though \eqref{eq:C-est-MAD} is the robust and preferred choice in practice \footnote{In the special situation where the innovations $\eta_k$ are independent and follow the same normal distribution, on can show that the estimate \eqref{eq:C-est-MAD} converges almost surely to \eqref{eq:C-est-mean} as the estimation window length $m$ goes to infinity (see, e.g., \cite{ref:Pham-Gia-2001})}. We use the notation $\smallO_m$ to denote and a sequence of random variables, indexed with $m$, which converges to zero in probability as $m$ goes to infinity.

We first note an expression of the innovation $\eta_k$ (see Appendix \ref{sec:appendex-A} for a derivation):
\begin{align} \label{eq:innov}
  \eta_k & = (1-K)^{k-1}(y_{0} - \hat x_{0}) + \sum_{i=1}^{k}(1-K)^{i-1}w_{k-i} + \sum_{i=1}^{k}(1-K)^{i-1} \delta v_{k-i},
\end{align}
where $\delta v_{k}$ is the difference of the measurement noise defined as:
\begin{align} \label{eq:def-delta-v}
   \delta v_{k} = v_{k+1} - v_{k}.
\end{align}
The first term in \eqref{eq:innov} vanishes with the choice of $\hat x_{0}$ in Algorithm 1, then we write \eqref{eq:innov} in a simpler form as
\begin{align} \label{eq:innov-in-two-terms}
  \eta_k & = \eta_{k,1} + \eta_{k,2}.
\end{align}
where we have defined
\begin{align}
\eta_{k,1} & = \sum_{i=1}^{k}(1-K)^{i-1} \delta v_{k-i} \label{eq:innov-first-part}, \\
\eta_{k,2} & = \sum_{i=1}^{k}(1-K)^{i-1}w_{k-i}. \label{eq:innov-second-part}
\end{align}
Consequently, one can put \eqref{eq:C-est-mean} into,
\begin{align} \label{eq:C-est-approx}
  \hat C(k,m) = S_{11}(k,m) + S_{22}(k,m) + 2\cdot S_{12}(k,m),
\end{align}
where $S_{11}(k,m)=\frac{1}{m}\sum_{i=k-m}^{k} ( \eta_{i,1} - \frac{1}{m+1}\sum_{j=k-m}^k \eta_{j,1} )^2$ is a sample estimate of the variance of $\eta_{k,1}$, which can be shown to converge to $\frac{2}{2-K}R$ in probability as the window length $m$ goes to infinity under the assumption in \eqref{eq:meas-noise} (see Appendix \ref{sec:appendex-B}); $S_{22}(k,m)=\frac{1}{m}\sum_{i=k-m}^{k} (\eta_{i,2} - \frac{1}{m+1}\sum_{j=k-m}^k \eta_{j,2})^2$ captures the variation of the signal as it is related only to $w_k$; and $S_{12}(k,m)=\frac{1}{m}\sum_{i=k-m}^{k} ( \eta_{i,1} - \frac{1}{m+1}\sum_{j=k-m}^k \eta_{j,1} )\cdot ( \eta_{i,2} - \frac{1}{m+1}\sum_{j=k-m}^k \eta_{j,2} )$, which reflects some correlation between the change of the signal and that of the measurement noise.

To simplify further analysis, we choose $K$ close to 1 so that we might only take the first term of the summations in \eqref{eq:innov-first-part} and \eqref{eq:innov-second-part}:
\begin{align}
\eta_{k,1} \approx \delta v_{k-1}, \quad \eta_{k,2} &\approx w_{k-1}. \label{eq:innov-sep-part-approx}
\end{align}
After some slightly involved but straightforward steps, one have $
    S_{12}(k,m) = \frac{1}{m}\sum_{i=k-m}^{k}(v_i-v_{i-1})w_{i-1} - \frac{1}{m}\left(\sum_{i=k-m}^{k}(v_i-v_{i-1})\right)\left(\frac{1}{m}\sum_{i=k-m}^{k}w_{i-1}\right)$,
which, under the assumption \eqref{eq:meas-noise}, converges to 0 in probability as $m$ goes large, if $\frac{1}{m}\sum_{i=k-m}^{k}w_{i-1}$ is finite (which is always the case in practice) and
\begin{align} \label{eq:uncorrected-w-v}
    \frac{1}{m}\sum_{i=k-m}^{k}(v_i-v_{i-1})w_{i-1} = \smallO_m.
\end{align}
In summary, with the assumptions in \eqref{eq:meas-noise} and \eqref{eq:uncorrected-w-v} and the approximations in \eqref{eq:innov-sep-part-approx}, we have from \eqref{eq:C-est-approx} that
\begin{align} \label{eq:C-est-further-approx}
  \hat C(k,m) \approx \frac{2}{2-K}R + \frac{1}{m}\sum_{i=k-m}^{k} \left( w_{i-1} - \frac{1}{m+1}\sum_{j=k-m}^k w_{j-1} \right)^2 + \smallO_m.
\end{align}
This means that estimation error defined at the beginning of this section can be approximated as
\begin{align} \label{eq:err-C-est-approx}
  \epsilon(k,m) \approx \frac{2-K}{2m}\sum_{i=k-m}^{k} \left( w_{i-1} - \frac{1}{m+1}\sum_{j=k-m}^k w_{j-1} \right)^2 + \smallO_m.
\end{align}
In particular, if the assumption \eqref{eq:process-noise-mean-var} on $w_k$ holds, then we have $\epsilon(k,m) \approx \frac{2-K}{2}Q + \smallO_m$ from \eqref{eq:err-C-est-approx}.
Therefore, we see that the estimate rendered by Algorithm 1 overestimates the variance of measurement noise if the signal is time-varying and the error depends on the variation of the signal in the moving estimation window. It can be also seen that the error is reduced by pushing the estimator gain $K$ close to 1 in view of the stability constraint $K\in(0,1)$ and the pre-requisite for the approximation in \eqref{eq:innov-sep-part-approx}. It is worth mentioning that the knowledge of the variation of the signal, if available, can certainly be used to reduce or even completely remove the first part of the error in \eqref{eq:err-C-est-approx}.

\section{Simulation results} \label{sec:sim}
\begin{figure}[!htbp]
\centering
\includegraphics[width=0.9\textwidth]{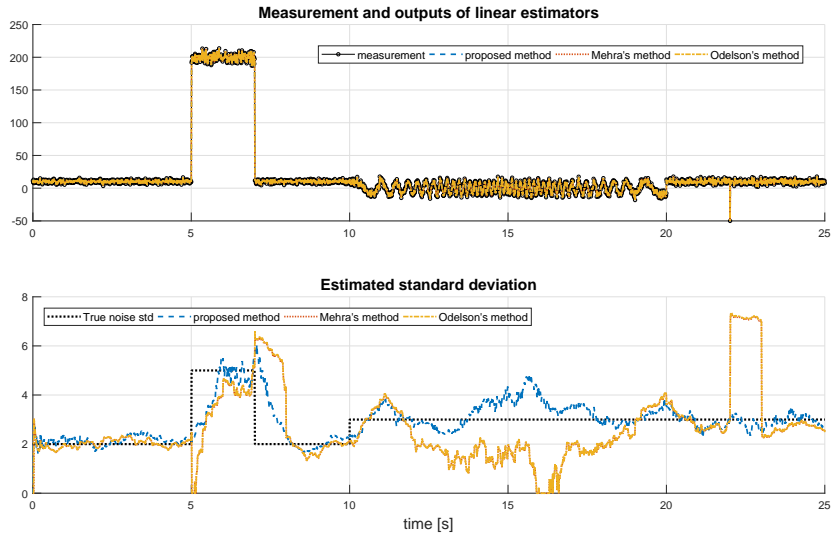}
\caption{Comparison of estimated measurement noise on a synthetic signal: Added noise is white with zero mean. The noise has piece-wise constant variance that is shown in the figure.}
\label{fig:sim-1}
\end{figure}

We compare the proposed method with the method described in \cite{ref:Mehra-1970} (with the key equations (22) and (23) therein) and the one described in \cite{ref:Odelson-2006}, which belong to the same class of correlation based methods as the proposed method in this paper (see, e.g., \cite{ref:Dunik-2009}). As in the comparison study \cite{ref:Dunik-2009}, auto-correlations (of the innovation sequence) up to lag 4 are used for both the methods. We use a synthetic signal with sampling frequency 100Hz to test each method. The synthetic signal can be seen as mimicking a signal measured from a real industrial process. The process is in steady-state at the start ($t=0$) and experiences some disturbance or unknown input at $t = 5$ seconds, which raises the magnitude of the signal; the process turns unstable with oscillating signal at about $t = 10$ seconds, the frequency of the oscillation increases first before decreases back to its original value; the process is stabilized (by, for example, an operator in practice) at time $t=20$ seconds and the signal become constant again with a outlier in measurement arriving at time $t=22$ seconds. The measurement noise at different time steps are independent and they all follow zero-mean normal distributions, the standard deviations of the distributions are shown in Figure \ref{fig:sim-1}.

To have a fair comparison, we use the same value for the estimator gain ($K=0.9902$) in all the three methods (steady-state Kalman filter gain in \cite{ref:Mehra-1970}); the same estimation window length of 100 samples is used for all three methods in estimating the variance of prediction errors. In addition, we take the initial value of the estimated state and its variance as $\hat x_0 = y_0$.

We note the following points from the simulation: (1) Due to the use of the same estimator gain, the output of all the three estimators coincide with each other. (2) The proposed method is able to estimate the varying variance of the noise reasonably well throughout the whole time period. It slightly overestimates the noise variance when the signal has fast oscillations, which agrees with the error analysis in Section \ref{sec:analysis}. (3) There are three places where both the methods in \cite{ref:Mehra-1970} and \cite{ref:Odelson-2006} go wrong. The first happens within the time period $5\leq t \leq 10$ where the signal has a couple of large jumps; the second occurs when the oscillation goes very fast; and the third happens with the presence of the outlier. The root cause of the issues is that both the methods uses a least-squares fitting to find the ``best" noise variance that matches variance of the innovation, which turns large when the three events occurs. These changes of magnitudes of innovation, however, are not related to the change of variance of the measurement noise.

\section{Conclusions}\label{sec:conclusions}
In this work, we have presented a simple and robust approach to estimate the variance of measurement noise. It belongs to the class of correlation based methods that were originated in 1970s and further developed in the past decades. The key to the robustness characteristic of our proposed approach is the use of a robust measure of the variability (i.e. the median absolute deviation) of the error between measurement and the predicted value of it from a linear time-invariant estimator, which is linked to the variance of the measurement noise by a simple explicit formula for 1-D signals. The estimate of the noise variance of our method is biased if the signal is time-varying and the bias increases with the variability of the signal. This is the cost of our assumption that the dynamics of the signal is completely unknown, which on the other hand makes the proposed approach widely applicable in practice. We hope this work can attract further research on robust extension of the popular correlation based methods for estimation of noise variances.


\appendix
\section{Derivation of equation \eqref{eq:innov}} \label{sec:appendex-A}
By \eqref{eq:state-pred} and \eqref{eq:state-update}, we have, when $k\geq 1$, $\hat x_k = \hat x^-_k + K(y_k - \hat x^-_k) = \hat x_{k-1} + K(y_k - \hat x_{k-1})$.
As a consequence, we have
\begin{subequations}\label{eq:yk-xk}
\begin{align}
  y_k - \hat x_k & = (1-K)(y_k - \hat x_{k-1}), \label{eq:yk-xk-0} \\
                 & = (1-K)(y_{k-1} - \hat x_{k-1} + w_{k-1} + \delta v_{k-1}), \label{eq:yk-xk-5}
\end{align}
\end{subequations}
where we have used the definition \eqref{eq:def-delta-v} in \eqref{eq:yk-xk-5}.
Going one time step back, one may obtain
\begin{align}\label{eq:yk-1-xk-1}
  y_{k-1} - \hat x_{k-1} & = (1-K)(y_{k-2} - \hat x_{k-2} + w_{k-2} + \delta v_{k-2}).
\end{align}
Combining \eqref{eq:yk-xk} and \eqref{eq:yk-1-xk-1}, we have $y_k - \hat x_k = (1-K)^2(y_{k-2} - \hat x_{k-2}) + (1-K)w_{k-1} + (1-K)^2w_{k-2} + (1-K)\delta v_{k-1} + (1-K)^2\delta v_{k-2}$.
Stepping further back in time, we end up with
\begin{align}\label{eq:yk-xk-recur}
  y_k - \hat x_k & = (1-K)^{k}(y_{0} - \hat x_{0}) + \sum_{i=1}^{k}(1-K)^iw_{k-i} + \sum_{i=1}^{k}(1-K)^i \delta v_{k-i}.
\end{align}
The innovation at time $k$, $k\geq 1$, can be written as
\begin{align} \label{eq:innov-tmp}
  \eta_k = y_k - \hat x^-_k = y_k - \hat x_{k-1} = \frac{1}{1-K}(y_k - \hat x_k).
\end{align}
where we have used \eqref{eq:yk-xk-0}. Combining \eqref{eq:innov-tmp} and \eqref{eq:yk-xk-recur}, one has \eqref{eq:innov}.

\section{Convergence of the first term on the right hand side of \eqref{eq:C-est-approx}} \label{sec:appendex-B}
It is clear that the first term on the right hand side of \eqref{eq:C-est-approx} is the unbiased sample variance of $\eta_{k,1}$ and thus converges to its variance in probability as $m$ goes large. By \eqref{eq:innov-first-part}, $\eta_{k,1} = \sum_{i=1}^{k}(1-K)^{i-1} \delta v_{k-i}$. Using the definition \eqref{eq:def-delta-v}, with a little effort, one can derive an alternative form of $\eta_{k,1}$ as $\eta_{k,1}= v_k - K\sum_{i=1}^{k-1}(1-K)^{k-1-i} v_i  -(1-K)^{k-1}v_{0}$. Then, by the assumption on the measurement noise \eqref{eq:meas-noise}, one easily checks that the variance of $\eta_{k,1}$ approaches $\left(1 + \frac{K^2}{1-(1-K)^2}\right)R = \frac{2}{2-K} R$ as $k\to\infty$.
Therefore, we have the first term on the right hand side of \eqref{eq:C-est-approx} converges to $\frac{2}{2-K} R$ in probability as $m$ and $k\to\infty$.

\bibliographystyle{plain}        
\bibliography{bibFile_noiseVariance}



\end{document}